\documentclass[12pt,eadjoint tfnpsf]{article}

\catcode`\@=11
\@addtoreset{equation}{section}

\global\arraycolsep=1pt

\usepackage{amsbsy,amssymb,latexsym,amsfonts, amsmath}
\usepackage{mathrsfs}
\usepackage{graphicx}

\RequirePackage[dvips,usenames]{color}
\definecolor{fireblick}{rgb}{0.698039,0.133333,0.133333}

\newcommand{\beq}{\begin{equation}}
\newcommand{\eeq}{\end{equation}}
\newcommand{\bea}{\begin{eqnarray}}
\newcommand{\eea}{\end{eqnarray}}

\newcommand{\CF}{{\mathcal F}}

\newcommand{\CN}{{\mathcal N}}
\newcommand{\CO}{{\mathcal O}}

\def\Tr{\mathop{\rm Tr}}
\newcommand\tr{\mathrm{tr}}

\renewcommand{\thefootnote}{\fnsymbol{footnote}}

\begin{document}
%
%
\begin{titlepage}

\begin{flushright}
\normalsize
~~~~
YITP-09-94\\
November, 2009 \\
\end{flushright}

\vspace{80pt}

\begin{center}
{\Large Penner Type Matrix Model and Seiberg-Witten Theory}\\
\end{center}

\vspace{25pt}

\begin{center}
{
Tohru Eguchi
and
Kazunobu Maruyoshi 
}\\
%
\vspace{15pt}
%
\it Yukawa Institute for Theoretical Physics, Kyoto University, Kyoto 606-8502, Japan\\
\end{center}
%
\vspace{20pt}
\begin{center}
Abstract\\
\end{center}
We discuss the Penner type matrix model recently proposed by Dijkgraaf and Vafa 
for a possible explanation of the relation between four-dimensional gauge theory and Liouville theory
by making use of the connection of the matrix model to two-dimensional CFT. 
We first consider the relation of gauge couplings defined in UV and IR regimes of $N_f=4$, 
$\CN = 2$ supersymmetric gauge theory being related as 
$q_{{\rm UV}}={\vartheta_2(q_{{\rm IR}})^4/\vartheta_3(q_{{\rm IR}})^4}$. 
We then use this relation to discuss the action of modular transformation on the matrix model 
and determine its spectral curve. 

We also discuss the decoupling of massive flavors from the $N_f=4$ matrix model 
and derive matrix models describing asymptotically free $\CN = 2$ gauge theories.  
We find that the Penner type matrix theory reproduces correctly 
the standard results of $\CN = 2$ supersymmetric gauge theories.


\vfill

\setcounter{footnote}{0}
\renewcommand{\thefootnote}{\arabic{footnote}}

\end{titlepage}

\section{Introduction}
\label{sec:intro}
  Recently an interesting observation has been made on the relationship 
  between the $\CN=2$ supersymmetric gauge theory and the 2d conformal field theory.
  In \cite{AGT}, it has been shown that Nekrasov's partition function \cite{Nekrasov} 
  of $\CN=2$ $SU(2)$ superconformal gauge theory can be identified with the correlation function of 2d Liouville theory.
  This relation was further studied and checked from various directions \cite{AGTother, MMMcoupling}.
  For general discussions on S-duality in $\CN=2$ superconformal gauge theories, see \cite{Gaiotto, GMN, Gaiottoother}.
  
  In ref.~\cite{DV}, the following matrix model with a Penner like action
    \bea
    W(M)
     =     \sum_{i = 1}^3 \mu_i \log (M - q_i), \hskip2cm q_1=0,\, q_2=1, \, 
     q_3=q
           \label{action}
    \eea
  has been proposed to describe $\CN=2$ $SU(2)$ superconformal gauge theory with four hypermultiplets.
  It is suggested that this matrix theory explains the correspondence of \cite{AGT},
  by making use of the CFT description of the matrix model \cite{MMM, KMMMP}.
  (This matrix model corresponds to $c = 1$ (or $b = i$) case in the Liouville theory.)
  The four mass parameters of the gauge theory are identified with $\mu_i$ in the matrix model action
  and $\mu_0$ which corresponds to the charge at infinity.
  These parameters satisfy the following relation \cite{DV}
    \bea
    \sum_{i = 1}^3 \mu_i + \mu_0
     =   - 2 g_s N,
           \label{massrelation}
    \eea
  where $N$ is the size of the matrix.
  
  We study this matrix model in details in this article 
  and show that they in fact correctly reproduce the physics of $SU(2)$ Seiberg-Witten theory 
  \cite{SeibergWitten, SeibergWitten2}.
  
  In section \ref{sec:gauge} 
  we first consider the M-theory curve for the $N_f=4$, $\CN=2$ supersymmetric $SU(2)$ gauge theory 
  based on the brane construction \cite{Witten}. 
  We study its period integrals and by comparing with those of the standard Seiberg-Witten curve 
  we obtain the relation between gauge coupling constants 
  at UV regime and IR regime as $q_{{\rm UV}}=\vartheta_2(q_{{\rm IR}})^4/\vartheta_3(q_{{\rm IR}})^4$. 
  The parameter $q$ which appears in the matrix model action (\ref{action}) is identified 
  as the UV coupling $q=q_{{\rm UV}}$.
  
  We discuss in section \ref{sec:matrix} the modular property of the matrix model (\ref{action}) 
  by making use of the IR-UV relation of gauge coupling constants. 
  We study the modular transformation of the spectral curve and determine its precise mass-dependence 
  by imposing modular invariance.
  
  In sections \ref{sec:asymptoticfree}, we consider the decoupling of massive flavors 
  and find matrix model actions corresponding to asymptotically free gauge theories with $N_f = 2, 3$. 
  We check that the discriminants of the spectral curve and free energy of these matrix models match well 
  with those of Seiberg-Witten theories.

\section{$\CN=2$ gauge theory and UV and IR gauge coupling constants}
\label{sec:gauge}

\subsection{M-theory curve of $N_f=4$, $SU(2)$ gauge theory}
\label{subsec:SWcurve}
  $\CN=2$ $SU(2)$ gauge theory with four hypermultiplets is known to be scale invariant
  and has the exactly marginal coupling constant
    \bea
    \tau_{{\rm UV}}
     =     \frac{\theta_{{\rm UV}}}{\pi} + \frac{8 \pi i}{g^2_{{\rm UV}}}.
           \label{gaugecouplingUV}
    \eea
  The flavor symmetry of the theory is $SO(8)$ whose maximal subgroup is $SU(2)^4$.
  While introducing the masses of the hypermultiplets breaks the conformal invariance, 
  the theory is modular invariant involving the triality of $SO(8)$ 
  which rotates the mass parameters \cite{SeibergWitten2}.
  We introduce the mass parameters of four hypermultiplets $m_{\pm}$ and $\tilde{m}_{\pm}$.
  By combining these, we define
    \bea
    m_0
     =     \frac{1}{2} (m_+ - m_-),~~
    m_2
     =     \frac{1}{2} (m_+ + m_-), ~~
    m_1
     =     \frac{1}{2} (\tilde{m}_+ - \tilde{m}_-), ~~
    m_3
     =     \frac{1}{2} (\tilde{m}_+ + \tilde{m}_-),
    \eea
  each of which is a mass parameter associated with each $SU(2)$ flavor symmetry.
  
  This gauge theory can be obtained by considering the intersecting brane system in type IIA string theory \cite{Witten}.
  The $SU(2)$ gauge part is induced by two D4-branes suspended between two NS5-branes.
  The D4-branes occupy the $x^{0, 1,2,3}$ and $x^6$ directions and 
  the NS5-branes occupy $x^{0, 1,2,3}$ and $x^{4,5}$ which are denoted by the complex coordinate $v$.
  A massive hypermultiplet can be introduced by a semi-infinite D4-brane attached to an NS5-brane.
  We here choose the configuration such that 
  two semi-infinite D4-branes are attached to the left NS5-brane and extended to $x^{6} = - \infty$
  and two more D4-branes are attached to the right NS5-brane extending 
  to $x^6 = + \infty$.
  By the M-theory uplift, the Seiberg-Witten curve \cite{SeibergWitten} of this theory
  becomes a hypersurface on $(t, v) \in \mathbb{C}^2$ where $t = e^{- (x^6 + i x^{10})/R}$:
    \bea
    (v - m_+) (v - m_-) t^2 + c_1 (v^2 + M v - U) t + c_2 (v - \tilde{m}_+) (v - \tilde{m}_-)
     =     0,
           \label{Mcurve}
    \eea
  where $M$ and $U$ are constants which depend on the masses and the Coulomb moduli $u$.
  The first and the third terms are determined as follows:
  in the large $t$, the first term is dominant and 
  we obtain $v \sim m_+, m_-$ which should be the masses of the hypermultiplets
  induced by the left semi-infinite D4-branes (large $t$ corresponds to $x^6 \rightarrow - \infty$).
  On the other hand, in the small $t$, the third term is dominant and $v \sim \tilde{m}_+, \tilde{m}_-$
  which are the mass parameters induced by the right semi-infinite D4-branes 
  (small $t$ corresponds to $x^6 \rightarrow + \infty$).
  The dimensional analysis and the regularity constraint in the massless limit show that
  $M$ is linear in the mass parameters and $U$ is linear in the Coulomb moduli parameter $u$ and also contains additional terms in mass squared.
  Also, $c_1$ and $c_2$ are the constants which parametrize the gauge coupling constant.
  Then, the curve can be written as
    \bea
    v^2 (t - 1)(t - q)
    &=&    v ((m_+ + m_-) t^2 + (1 + q)Mt + q(\tilde{m}_+ + \tilde{m}_-) )
           \nonumber \\
    & &   - m_+ m_- t^2 - (1 + q)Ut - q \tilde{m}_+ \tilde{m}_-,
    \eea
  where we have chosen that $c_1 = - (1 + q)$ and $c_2 = q$.
  
  By eliminating the terms linear in $v$ and changing the coordinate as $v = x t$, 
  the curve can be written as the following form \cite{Gaiotto}:
    \bea
    x^2
     =     \left( \frac{m_2 t^2 + (1 + q) \frac{M}{2} t + m_3 q}{t(t - 1)(t - q)} \right)^2
         + \frac{(m_0^2 - m_2^2) t^2 - (1 + q) U t + (m_1^2 - m_3^2) q}{t^2 (t -1) (t - q)},
           \label{SWcurve}
    \eea
  where $x^2 dt^2$ is considered as a quadratic differential on a sphere 
  ($t$ is a coordinate on the sphere)
  and has double poles at $t = 0, 1, q, \infty$.
  In this coordinate, the Seiberg-Witten one-form is 
    \bea
    \lambda_{{\rm SW}}
     =     \frac{x dt}{2 \sqrt{2} \pi i},
    \eea
  where we have divided by the factor $2 \sqrt{2} \pi i$ 
  in order to be consistent with the convention in later sections.
  Here, $(t, x)$ are local coordinates on the cotangent bundle of the sphere.
  The Seiberg-Witten curve is the double cover of this sphere with four punctures.
  
  The moduli space of the sphere with four punctures is parametrized by $q$ above.
  As discussed in \cite{Witten, Gaiotto}, this moduli space can be identified 
  with the parameter space of the exactly marginal operator of 4d gauge theory,
  that is the gauge coupling constant (\ref{gaugecouplingUV}).
  Therefore, we identify $q$ with $q_{{\rm UV}} = e^{\pi i \tau_{{\rm UV}}}$.
  Note that this is the UV gauge coupling constant.
  The effective gauge coupling at IR is derived using the full information of the Seiberg-Witten theory.
  We will discuss this point in the next subsection.
  
  The masses of the flavors can be read from the residues of $\lambda_{{\rm SW}}$ at the punctures.
  As seen in \cite{Gaiotto}, each puncture is associated with each $SU(2)$ flavor symmetry.
  Indeed, we can easily see that the residues at $t = 0, \infty$ are $\pm m_1$ and $\pm m_0$
  where the signs represent their value on the upper and the lower sheets respectively.
  (More precisely, the residues are $\pm \frac{m_{0, 1}}{2 \sqrt{2} \pi i}$, 
  but we will ignore the factor $2 \sqrt{2} \pi i$ in what follows when discussing the residues.)
  Also, the requirement that the residues at $t = 1, q_{{\rm UV}}$ should be $\pm m_2, \pm m_3$ determines the form of $M$ as follows:
    \bea
    M
     =   - \frac{2 q_{{\rm UV}}}{1 + q_{{\rm UV}}} (m_2 + m_3).
    \eea
  
  The Coulomb branch vev $a$ can be obtained by the period integral
    \bea
    a
     =     \oint_A \lambda_{{\rm SW}}.
    \eea
  This is the function of the Coulomb moduli $u$.
  Also, the dual $B$ cycle integral computes the derivative of the prepotential
    \bea
    a_D
     =     \frac{\partial \CF}{\partial a}
     =     \oint_B \lambda_{{\rm SW}}.
    \eea
  The moduli derivative of $\lambda_{{\rm SW}}$ is a holomorphic one-form 
    \bea
    \omega 
     =     \frac{\partial \lambda_{{\rm SW}}}{\partial u}
     =   - \frac{(1 + q_{{\rm UV}})}{4 \pi i \sqrt{2 P_4(t)}} \frac{\partial U}{\partial u} dt,
           \label{holomorphiconeform}
    \eea
  whose $A$ and $B$ periods give $\frac{\partial a}{\partial u}$ and $\frac{\partial a_D}{\partial u}$.
  $P_4(t)$ is a degree-four polynomial whose precise form does not concern us.
  As we will see in subsection \ref{subsec:coupling}, 
  $\frac{\partial U}{\partial u}$ becomes $1$ in the weak coupling limit: $q_{{\rm UV}} \rightarrow 0$.
  
  We now consider a way to extract the parameter $U$ (essentially $\langle \Tr \phi^2 \rangle$) 
  from the M-theory curve above.
  In the pure $\CN=2$ $SU(N)$ gauge theory, it is known that such an operation exists  \cite{DGKV, Gopakumar, NSWN=2}
    \bea
    \langle \Tr \phi^2 \rangle
     =     \frac{1}{2 \pi i} \oint v \lambda_{{\rm SW}}
     =     \frac{1}{2 \pi i} \oint x^2 t dt,
    \eea
  where $v$ is the coordinate in (\ref{Mcurve}).
  Also in the $SU(2)$ gauge theory with four massive flavors,
  we can use this kind of integral to extract the $U$ parameter.
  Actually, since the term including $U$ in $x^2$ is
    \bea
    - \frac{(1 + q_{{\rm UV}})U}{t (t - 1) (t - q_{{\rm UV}})},
    \eea
  the integral 
    \bea
    \frac{1}{2 \pi i} \oint_{C_{\infty}} x^2 t^n dt,
    ~~~~{{\rm for}}~n \geq 2
    \eea
  is linear in $U$ and contains terms involving mass parameters.
  We denote a contour around infinity in the counterclockwise direction as $C_{\infty}$.
  In the simplest case ($n = 2$), we easily obtain
    \bea
    \frac{1}{2 \pi i}\oint_{C_{\infty}} x^2 t^2 dt
     =     (1 + q_{{\rm UV}}) U - (1 + q_{{\rm UV}}) m_0^2
         + (q_{{\rm UV}} - 1) m_2^2 + 2 q_{{\rm UV}} m_2 m_3.
           \label{U}
    \eea
  This relation becomes an important point when we compare the gauge theory with the matrix model 
  in section \ref{sec:matrix}.

\subsection{UV and IR gauge coupling constants}
\label{subsec:coupling}
  The M-theory curve (\ref{SWcurve}) which we have seen in the previous subsection is different 
  from the standard Seiberg-Witten curve introduced in \cite{SeibergWitten2}.
  In this subsection, we will see that the comparison of the M-theory curve and the standard  Seiberg-Witten curve leads to 
  a relation between the UV and IR gauge coupling constants \cite{DKM, GKMW, AGT, MMMcoupling}. We first consider the massless case for simplicity. 

On the one hand, we have the standard form of the Seiberg-Witten curve \cite{SeibergWitten2}
    \bea
    y^2
     =     4 x^3 - g_2 u^2 x - g_3 u^3,
           \label{SWcurveoriginal}
    \eea
  where $u$ parametrizes the Coulomb moduli and is related to $\langle \Tr \phi^2 \rangle$.
  The IR gauge coupling constant 
    \bea
    \tau_{{\rm IR}}
     \equiv
           \frac{\theta_{{\rm IR}}}{\pi} + \frac{8 \pi i}{g_{{\rm IR}}^2}
    \eea
  is calculated from the period integrals of the holomorphic one-form $\omega$
  which is defined as 
    \bea
    \omega
     =     \frac{\sqrt{2}}{4 \pi} \frac{d x}{y}
     =     \frac{\sqrt{2}}{4 \pi} \frac{d x}{\sqrt{4 x^3 - g_2 u^2 x - g_3 u^3}},
           \label{holo}
    \eea
  where $g_2$ and $g_3$ are the functions of $q_{{\rm IR}} = e^{\pi i \tau_{{\rm IR}}}$:
    \bea
    g_2(\omega_1, q_{{\rm IR}})
    &=&    \left( \frac{\pi}{\omega_1} \right)^4 \frac{1}{24} 
           \left( \vartheta_3(q_{{\rm IR}})^8 + \vartheta_2(q_{{\rm IR}})^8 + \vartheta_4(q_{{\rm IR}})^8 \right),
           \nonumber \\
    g_3(\omega_1, q_{{\rm IR}})
    &=&    \left( \frac{\pi}{\omega_1} \right)^6 \frac{1}{432} 
           \left( \vartheta_4(q_{{\rm IR}})^4 - \vartheta_2(q_{{\rm IR}})^4 \right)
           \left( 2 \vartheta_3(q_{{\rm IR}})^8 + \vartheta_4(q_{{\rm IR}})^4 \vartheta_2(q_{{\rm IR}})^4 \right),
    \eea
  and $2 \omega_1 = \pi$ in this case.
    $A$-cycle integral of $\omega$ gives
    \bea
    \frac{\partial a}{\partial u}
     =     \frac{\sqrt{2}}{4 \pi} \oint_A \frac{d x}{\sqrt{4 x^3 - g_2 u^2 x - g_3 u^3}}
     =     \frac{1}{2 \sqrt{2u}},
    \eea
  which leads to $a = \frac{1}{2} \sqrt{2u}$.

 As pointed out in \cite{DKM}, however, the gauge coupling constant receives the correction due to instanton effects even in $N_f=4$ theory 
  and, therefore, $q_{{\rm IR}} \neq q_{{\rm UV}}$.
  As we saw in the previous subsection, 
  we obtain the following curve in the massless limit of the M-theory curve:
    \bea
    x^2
     =   - \frac{(1 + q_{{\rm UV}}) U}{t (t - 1)(t - q_{{\rm UV}})},
           \label{SWcurve'}
    \eea
  where $q_{{\rm UV}}$ is the UV gauge coupling constant.
  Note that the parameter $U$ in the above curve would be different from $u$ in the curve (\ref{SWcurveoriginal}).
  Dimensional analysis shows that $U$ is proportional to $u$:
    \bea
    U 
     =     A u,
    \eea
  where $A$ depends only on the UV gauge coupling $q_{{\rm UV}}$.
  The holomorphic one-form is given by
    \bea
    \omega
     =     \frac{1}{4 \sqrt{2} \pi i} 
           \sqrt{\frac{- (1 + q_{{\rm UV}})A}{u}} \frac{dt}{\sqrt{t (t - 1) (t - q_{{\rm UV}})}}.
    \eea
  By changing the coordinate as $t \rightarrow \tilde{t} + \frac{1 + q_{{\rm UV}}}{3}$
  and by rescaling as $\tilde{t} \rightarrow 4 z$, 
  $t (t - 1) (t - q_{{\rm UV}})$ is transformed to
    \bea
    16 \left( 4z^3 - \frac{1}{12} (1 - q_{{\rm UV}} + q_{{\rm UV}}^2) z
    - \frac{1}{432} (2 - 3q_{{\rm UV}} - 3q_{{\rm UV}}^2 + 2 q_{{\rm UV}}^3) \right)
     \equiv
           16 y^2_0.
    \eea
  Then, the holomorphic one-form reads as
    \bea
    \omega
     =     \frac{1}{4 \sqrt{2} \pi} \sqrt{\frac{(1 + q_{{\rm UV}}) A}{u}}
           \frac{dz}{y_0}.
    \eea
    
  Now, we want to compare this with the one-form (\ref{holo}).
  In order to reproduce the $A$ period $\frac{1}{2 \sqrt{2u}}$, we should obtain
    \bea
    \oint_A \frac{dz}{y_0}
     =     \frac{2 \pi}{\sqrt{(1 + q_{{\rm UV}}) A}}.
    \eea
  Therefore, the following relations must be satisfied:
    \bea
    \frac{1}{12} (1 - q_{{\rm UV}} + q_{{\rm UV}}^2)
    &=&    g_2(\omega_1, q_{{\rm IR}}),
           \nonumber \\
    \frac{1}{432} (2 - 3q_{{\rm UV}} - 3q_{{\rm UV}}^2 + 2 q_{{\rm UV}}^3)
    &=&    g_3(\omega_1, q_{{\rm IR}}),
           \label{relations}
    \eea
  and the period is given by
    \bea
    2 \omega_1
     =     \frac{2 \pi}{\sqrt{(1 + q_{{\rm UV}}) A}}.
    \eea
  By using formula $\vartheta_4^4 = \vartheta_3^4 - \vartheta_2^4$, the right hand sides of (\ref{relations}) 
  can be calculated as
    \bea
    g_2
    &=&    \frac{1}{12} (1 + q_{{\rm UV}})^2 A^2 \vartheta_3^8
           \left( 1 - \frac{\vartheta_2^4}{\vartheta_3^4}
         + \left( \frac{\vartheta_2^4}{\vartheta_3^4} \right)^2 \right),
           \nonumber \\
    g_3
    &=&    \frac{1}{432} (1 + q_{{\rm UV}})^3 A^3 \vartheta_3^{12}
           \left( 2 - 3 \frac{\vartheta_2^4}{\vartheta_3^4} - 3 \left( \frac{\vartheta_2^4}{\vartheta_3^4} \right)^2 
       + 2 \left( \frac{\vartheta_2^4}{\vartheta_3^4} \right)^3 \right).
    \eea
  This shows that
    \bea
    q_{{\rm UV}}
     =     \frac{\vartheta_2(q_{{\rm IR}})^4}{\vartheta_3(q_{{\rm IR}})^4}
     =     16 q_{{\rm IR}} - 128 q_{{\rm IR}}^2 + 704 q_{{\rm IR}}^3 - 3072 q_{{\rm IR}}^4 + \ldots.
           \label{UVIRrelation}
    \eea
  and 
    \bea
    A
     =     \frac{1}{\vartheta_2(q_{{\rm IR}})^4 + \vartheta_3(q_{{\rm IR}})^4},
           \label{a}
    \eea
  which implies that $U = u + \CO(q_{{\rm UV}}) = \langle \Tr \phi^2 \rangle + \CO(q_{{\rm UV}})$
  in the weak coupling limit.
  The relation (\ref{UVIRrelation}) between the UV and IR gauge coupling constants has already been suggested
  in \cite{GKMW} from the topological string analysis and derived in \cite{MMMcoupling} 
  along the similar line as above.
  
  Let us next consider the case with the massive flavors.
  In this case, the parameter $U$ appeared in the M-theory curve (\ref{SWcurve}) receives
  additional contributions due to mass terms.
  However, the form of $A$ (\ref{a}) should not change.
  Therefore, we obtain
    \bea
    \frac{\partial U}{\partial u}
     =     \frac{1}{\vartheta_2^4 + \vartheta_3^4}.
    \eea
  It follows from this that the holomorphic one-form (\ref{holomorphiconeform}) becomes
    \bea
    \omega 
     =   - \frac{dt}{4 \pi i \vartheta_3^4 \sqrt{2 P_4(t)}}.
    \eea
  The $A$-cycle integral of this one-form can be translated into the following standard form 
  of the elliptic integral of the first kind, by the coordinate transformation:
    \bea
    \int_0^1 \frac{d u}{\sqrt{(1 - u^2)(1 - k^2 u^2)}},
    \label{standard}
    \eea
  up to a constant factor, where $k$ is a function determined from $P_4(t)$.
  We also obtain the similar form for the $B$ cycle integral.
  The IR gauge coupling constant is given by the period of this one form:
  $\tau_{{\rm IR}} = \frac{\oint_B \omega}{\oint_A \omega}$ and this in turn leads to
    \bea
    k^2
     =     \frac{\vartheta_2(q_{{\rm IR}})^4}{\vartheta_3(q_{{\rm IR}})^4}.
    \eea
  This determines the IR gauge coupling constant in terms of the UV one.

\section{Matrix model and modular invariance}
\label{sec:matrix}
  In this section, we analyze the matrix model with the action (\ref{action}) which was proposed to describe
  $\CN=2$ $SU(2)$ gauge theory with four massive hypermultiplets and study its modular properties.
  We first review the general technique of the one matrix model
  and then consider the modular invariance of the spectral curve.
  
  We define the free energy of the matrix model as follows:
    \bea
    \exp \left( \frac{F_{m}}{g_s^2} \right)
    &=&    \int d M \exp \left( \frac{1}{g_s} W(M) \right)
           \nonumber \\
    &=&    \int \left( \prod_{I = 1}^N d \lambda_I \right)
           \exp \left[ \frac{1}{g_s} \left( \sum_{I = 1}^N W(\lambda_I) + g_s \sum_{I<J} \log (\lambda_I - \lambda_J)^2 
           \right) \right],
    \eea
  where $M$ is the hermitian matrix whose size is $N$ and $W(M)$ is (\ref{action}).
  In the second line, we have switched to integrals over the eigenvalues $\lambda_I$.
  The last term comes from the Vandermonde determinant.
  The critical points are determined by the equation of motion
    \bea
    \sum_{i = 1}^3 \frac{\mu_i}{\lambda_I - q_i}
         + 2 g_s \sum_{J (\neq I)} \frac{1}{\lambda_I - \lambda_J}
     =     0.
           \label{eom}
    \eea
  By ignoring the second term, we obtain two critical points $e_p$ ($p = 1, 2$).
  The positions $q_i$ will be chosen as $q_1 = 0, q_2 = 1$ and $q_3 = q$.
  Let each $N_p$ ($p = 1, 2$) be the number of the eigenvalues which are classically at $e_p$.
  Let us define the resolvent of the matrix model as
    \bea
    R_{m}(z)
     =     g_s \tr \frac{1}{z - M}.
    \eea
  
  In what follows, we take the large $N$ limit while $\mu_p$ and the filling fractions $\nu_i \equiv g_s N_p$ are fixed.
  The loop equation in the large $N$ limit is written as
    \bea
    \langle R_m (z) \rangle^2
     =   - \langle R_{m} (z) \rangle W'(z) + \frac{f(z)}{4},
    \eea
  where 
    \bea
    f(z) 
     \equiv
           4 g_s \tr \left< \frac{W'(z) - W'(M)}{z - M}  \right>
     =     \sum_{i = 1}^3 \frac{c_i}{z - q_i}
           \label{ci}
    \eea
  has simple poles at $z = q_i$.
  Note that this function is not a polynomial in contrast to the ordinary case of polynomial action.
  $c_i$ are functions of $\mu_i$ and the filling fractions but they satisfy $\sum_i c_i = 0$
  which follows from the equations of motion (\ref{eom}).
  
  Let us define the meromorphic one-form $\lambda_m = \frac{x(z) dz}{2 \sqrt{2} \pi i}$ such that
    \bea
    x(z)^2
     \equiv     
           \left( 2 \langle R_m(z) \rangle + W'(z) \right)^2
     =     W'(z)^2 + f(z).
           \label{spectral}
    \eea
  This one-form has simple poles at $z = q_i, \infty$ with the residues $\mu_i, \mu_0$.
  (As in the field theory analysis, the residues are precisely $\frac{\mu_i}{2 \sqrt{2} \pi i}$ 
  by the definition of $\lambda_{m}$.
  But, for convenience, we will ignore the factor in the denominator when discussing the residues.)
  Note that the residue of the pole at $z = \infty$ can be evaluated 
  by observing $\langle R_m \rangle \sim \frac{g_s N}{z}$ and $W'(z) \sim \frac{\sum_{i = 1}^3 \mu_i}{z}$
  at large $z$ and by using the relation (\ref{massrelation}).
  
  The spectral curve of the matrix model (\ref{spectral}) looks like the M-theory curve.
  The former includes the parameters $\mu_i$ which will be identified with the masses $m_i$ of the latter.
  These are the residues of the meromorphic one-form $\lambda_m$ and $\lambda_{{\rm SW}}$ at the simple poles.
  Hence, the identification $q = q_{{\rm UV}}$ is needed.
  To compare more precisely, we should add a dimensionful parameter by $g_s \rightarrow g_s \epsilon$
  since the parameters in the matrix model are dimensionless.
  In the analysis of the gauge theory, the parameter $U$ is obtained from $\oint x^2 t^n dt$ for $n \geq 2$ (\ref{U}).
  In the matrix model language, the similar integral gives
    \bea
    \frac{1}{2 \pi i } \oint_{C_{\infty}} x^2 z^2 dz
     =     \frac{1}{2 \pi i} \oint_{C_\infty} (2 R_m + W')^2 z^2 dz 
     =     4 \mu_0 g_s \langle \tr M \rangle + 2 \mu_0 (\mu_2 + q \mu_3),
           \label{matrixU}
    \eea
  where $C_{\infty}$ is the contour around infinity.
  Therefore, the precise correspondence between the matrix model and the gauge theory is given 
  by the following identification:
    \bea
    & &    4 \mu_0 g_s \langle \tr M \rangle + 2 \mu_0 (\mu_2 + q \mu_3)
           \nonumber \\
    & &    ~~~~~~~~~~
     =     (1 + q_{{\rm UV}}) U - (1 + q_{{\rm UV}}) m_0^2
         + (q_{{\rm UV}} - 1) m_2^2 + 2 q_{{\rm UV}} m_2 m_3,
           \label{identification4}
    \eea
  and also $\mu_i = m_i$, $q = q_{{\rm UV}}$.
  
  We can rewrite $\langle \tr M \rangle$ in terms of the coefficients $c_i$ in (\ref{ci}).
  First of all, the residue at $z = \infty$ imposes the following constraint on $c_i$:
    \bea
    c_2 + q c_3
     =     \mu_0^2 - \left( \sum_{i = 1}^3 \mu_i \right)^2.
           \label{crelation}
    \eea
  Therefore, by recalling $\sum_i c_i = 0$, only one of ${c_i}'s$ is independent.
  By using (\ref{crelation}) and $\sum c_i = 0$, the spectral curve can be written as 
    \bea
    x^2
     =     \left( \frac{\mu_1}{z} + \frac{\mu_2}{z - 1} + \frac{\mu_3}{z - q} \right)^2
         + \frac{(\mu_0^2 - \left( \sum_{i = 1}^3 \mu_i \right)^2)z + q c_1}{z (z - 1)(z - q)}.
           \label{spectral2}
    \eea
  Also, the filling fractions can be obtained by the A-cycle integrals of the one-form:
    \bea
    \frac{\nu_p}{\sqrt{2} \pi i} 
     =     \frac{1}{2 \pi i} \oint_{A_p} \lambda_m,
           ~~~
           p = 1, 2,
    \eea
  where $A_p$ are the cycles around the branch cuts (corresponding to two critical points).
  From this, we can in principle determine $\nu_p$ as the function of $\mu_i$ and $c_1$.
  In other words, $c_1$ can be treated as an independent parameter.

  It follows from the dimensional analysis that $c_1$ has mass dimension two. 
  Therefore, it is natural to consider that $c_1$ is a linear function in $U$ of the 
  M-theory curve.
  Indeed, by substituting the explicit expression (\ref{spectral2}) into the left hand side of (\ref{matrixU}),
  we obtain
    \bea
    \frac{1}{2 \pi i } \oint_{C_{\infty}} x^2 z^2 dz
    &=&  - q c_1 - (1 + q) \mu_0^2 + (1 + q) \mu_1^2 + (q - 1) \mu_2^2
           \nonumber \\
    & &   + (1 - q) \mu_3^2 + 2 q \mu_1 \mu_2 + 2 \mu_3 \mu_1.
    \eea
  This and (\ref{identification4}) lead to
    \bea
    q_{{\rm UV}} c_1
    &=&    (1 + q_{{\rm UV}}) m_1^2 + (1 - q_{{\rm UV}}) m_3^2 + 2 q_{{\rm UV}} m_1 m_2 
           \nonumber \\
    & &    ~~~~~
         - 2 q_{{\rm UV}} m_2 m_3 + 2 m_3 m_1 - (1 + q_{{\rm UV}}) U.
           \label{identification}
    \eea
  We can show that under this relation, $\mu_i = m_i$ and $q = q_{{\rm UV}}$,
  the meromorphic one-form $\lambda_m$ is equal to $\lambda_{{\rm SW}}$.
  Therefore, the $A$ and $B$-periods of $\lambda_m$ and $\lambda_{SW}$ coincide, e.g., $\sqrt{2} \nu_1(u) = a(u)$.
  
  Also, $u$-derivative of $\lambda_m$ gives a holomorphic one-form:
    \bea
    \omega_m
     =     \frac{\partial \lambda_m}{\partial u}
     =   - \frac{dz}{4 \pi i \vartheta_3^4 \sqrt{2 P_{m4}(z)}},
    \eea
  where $P_{m4}(z)$ is the same degree 4 polynomial as $P_4$ in (\ref{holomorphiconeform}).
  The effective gauge coupling constant $\tau_{{\rm IR}}$ can be obtained from the period of $\omega_m$.
  
\subsection{Modular invariance}
\label{subsec:modular}
  It is known that the standard Seiberg-Witten curve of $SU(2)$ gauge theory with $N_f = 4$ is 
  invariant under modular transformation in \cite{SeibergWitten2}.
  We will see in this subsection that the spectral curve (\ref{spectral2})
  which was identified with the M-theory curve (\ref{SWcurve}) can be made modular invariant.
  
  To begin with, we consider the massless limit of the curve (\ref{spectral2})
    \begin{eqnarray}
    x^2
     =   - {(1+q_{{\rm UV}})U\over z(z-1)(z-q)}
     =   - {\displaystyle{{u\over \vartheta_3^4}}\over z(z-1)(z-q)}.
    \end{eqnarray} 
    where we have used the relation (\ref{a}).  
  This is invariant under the following transformations
    \begin{eqnarray}
    I:(z, x) \rightarrow (1 - z, x), && \hskip3mm q_{{\rm UV}} \rightarrow 1 - q_{{\rm UV}},
    \hskip3mm u \rightarrow -u, \hskip3mm S \\
    II:(z, x) \rightarrow (\frac{1}{z}, - z^2 x), && \hskip3mm q_{{\rm UV}} \rightarrow \frac{1}{q_{{\rm UV}}},\hskip10mm u\rightarrow u,\hskip4mm STS
    \end{eqnarray}
  Since $q_{{\rm UV}} = \vartheta_2^4/\vartheta_3^4$ as seen in subsection \ref{subsec:coupling},
  the former transformation is the exchange of 
  $\vartheta_2^4$ and $\vartheta_4^4$,
  which is the S-transformation.
  Also, the latter one can be seen as the STS-transformation:
  $\vartheta_2^4 \leftrightarrow \vartheta_3^4$.
  We note that the sign of $u$ changes under S-transformation. 
  This behavior is the same as in the case of standard SW curve. 
  In fact in (\ref{SWcurveoriginal}) $g_2$ and $g_3$ are even and odd and $u$ changes sign under S-transformation. 
  
  Next, let us consider the massive case.
  As analyzed in \cite{SeibergWitten2}, the modular transformation in this case involves the triality of $SO(8)$
  which permutes the $SU(2)$ flavor symmetries.
  In our notation, this permutes the mass parameters $m_i$ ($i = 0, \ldots, 3$) associated with these $SU(2)$'s.
  Under the S- and STS-transformations, positions and residues of the poles of $\lambda_m$ are transformed as 
    \begin{eqnarray}
    & &    I:(0, 1, q_{{\rm UV}}, \infty) \rightarrow (1, 0, 1-q_{{\rm UV}}, \infty),\hskip5mm 
           m_1 \leftrightarrow m_2,\\
    & &    II:(0,1,q_{{\rm UV}}, \infty) \rightarrow (\infty,1,{1\over q_{{\rm UV}}},0), \hskip8mm  
           m_0 \leftrightarrow m_1.
    \end{eqnarray}

  Under these transformations, the spectral curve should be invariant.
  Let us substitute (\ref{identification}) into (\ref{spectral2}) and we obtain
  \begin{eqnarray}
&&  x^2(z;m_i;q_{{\rm UV}})=\left({m_1 \over z}+{m_2\over z-1}+{m_3\over z-q_{{\rm UV}}}\right)^2 
    +{(m_0^2-(\sum_{i=1}^3 m_i)^2)z+(1+q_{{\rm UV}})m_1^2\over \null}  \\
&&  \hskip5cm {+(1-q_{{\rm UV}})m_3^2 +2q_{{\rm UV}}m_1m_2-2q_{{\rm UV}}m_2m_3+2m_1m_3-(1+q_{{\rm UV}})U\over z(z-1)(z-q_{{\rm UV}})}. 
\nonumber
  \end{eqnarray}
We then impose the conditions
 \begin{eqnarray}
 &&x^2(z;m_0,m_1,m_2,m_3;q_{{\rm UV}})=x^2(1-z;m_0,m_2,m_1,m_3;1-q_{{\rm UV}}),\\
 &&x^2(z;m_0,m_1,m_2,m_3;q_{{\rm UV}})={1\over z^4}x^2({1\over z};m_1,m_0,m_2,m_3:{1\over q_{{\rm UV}}}).
 \end{eqnarray}
Requirement of modular invariance determines completely the mass dependence of the parameter $U$. The solution to the above conditions is given by 
    \bea
    (1 + q_{{\rm UV}})U
     =     \frac{u}{\vartheta_3^4} - q_{{\rm UV}} (m_2 + m_3)^2
         + \frac{1 + q_{{\rm UV}}}{3} \left( \sum_{i = 0}^3 m_i^2 \right).
         \label{solution}
    \eea

\subsection{Relation for $\langle \tr M^m \rangle$}
\label{subsec:recursion}
  In (\ref{identification4}), we have written down the parameter $U$ in terms of $\langle \tr M \rangle$
  by using the integral $\oint x^2 z^2 dz$.
  As seen in subsection \ref{subsec:SWcurve}, we can in principle use the other integrals as
    \bea
    \frac{1}{2 \pi i} \oint_{C_{\infty}} x^2 z^n dz, 
    ~~~{\rm for}~n \geq 2,
    \eea
  in order to extract the parameter $U$.
  This in turn gives the relations for $\langle \tr M^m \rangle$'s.
  For example, we consider $n = 3$ case.
  From the M-theory curve, we can compute
    \bea
    \frac{1}{2 \pi i} \oint_{C_{\infty}} x^2 t^3 dt
    &=&    4 m_0 (1 + q_{{\rm UV}}) g_s \langle \tr M \rangle + q_{{\rm UV}} m_0^2 - q_{{\rm UV}} m_1^2 
         + (q_{{\rm UV}} - 1) m_2^2
           \nonumber \\
    & &  + q_{{\rm UV}} (1 - q_{{\rm UV}}) m_3^2 + 2 (1 + q_{{\rm UV}}) m_0 (m_2 + q_{{\rm UV}} m_3),
    \eea
  where we have used (\ref{identification4}).
  On the other hand, the same integral for the matrix model is calculated as
    \bea
    \frac{1}{2 \pi i} \oint_{C_{\infty}} x^2 z^3 dz
    &=&    4 \mu_0 g_s \langle \tr M^2 \rangle 
         - 4 g_s \langle \tr M \rangle (g_s \langle \tr M \rangle + m_2 + q_{{\rm UV}} m_3)
           \nonumber \\
    & &  - (m_2 + q_{{\rm UV}} m_3)^2 + 2 m_0 (m_2 + q_{{\rm UV}}^2 m_3).
    \eea
  From these equations, we obtain
    \bea
    4 \mu_0 g_s \langle \tr M^2 \rangle
    &=&    4 g_s \langle \tr M \rangle (g_s \langle \tr M \rangle + (1 + q_{{\rm UV}}) m_0 + m_2 + q_{{\rm UV}} m_3)
           \nonumber \\
    & &  + q_{{\rm UV}} (m_0^2 - m_1^2 + m_2^2 + m_3^2) + 2 q_{{\rm UV}} (m_2 m_3 + m_0 m_2 + m_0 m_3).
    \eea
  Similarly, $\langle \tr M^m \rangle$ for $m \geq 3$ can be written 
  in terms of the lower order ones.

\subsection{Free energy}
  So far, we have learned that M-theory curve and spectral curve of matrix model can be identified. 
  However, it is not so obvious that the matrix model free energy coincides with the prepotential of gauge theory,
  in other words, the B-period of $\lambda_m$ (differential of the matrix theory) 
  is written as $\frac{\partial F_m}{\partial \nu_1}$.
  Let us see this below.
  We consider the free energy and its derivative with respect to the filling fractions.
  First of all, we rewrite the free energy as
    \bea
    e^{F_m/g_s^2}
    &=&    \exp \left[ \frac{1}{g_s^2} \left( \int d \lambda \rho(\lambda) W(\lambda) 
         + \int d \lambda d \lambda' \rho(\lambda) \rho(\lambda') \log |\lambda - \lambda'| \right) \right],
    \eea
  where $\rho(\lambda)$ is the eigenvalue distribution function normalized as
    \bea
    \int d\lambda \rho(\lambda) 
     =     g_s N,
    \eea
  or $\rho(\lambda) = g_s \sum_I \delta(\lambda - \lambda_I)$.
  The resolvent can be written as
    \bea
    R_m(z)
     =     \int d\lambda \frac{\rho(\lambda)}{z - \lambda}.
    \eea
    
  Now, let us consider the derivative of the free energy with respect to the filling fractions $\nu_p$.
  Here, we will follow the discussion in \cite{NSW, CSW}.
  The variations with respect to $\nu_p$ can be considered as the shift 
  $\rho(\lambda) \rightarrow \rho(\lambda) + (\delta \nu_p) \delta(\lambda - e_p^+)$
  where $e_p^+$ are any points on the branch cuts (except for the end points of the cuts).
  Therefore, the free energy is shifted as
    \bea
    \delta F_m
    &=&    (\delta \nu_p) \left[ W(e_p^+) + 2 \int d\lambda \rho (\lambda) \log |\lambda - e_p^+| \right]
           \nonumber \\
    &=&    (\delta \nu_p) \left[ W(e_p^+) + \frac{2}{2 \pi i} \oint_{A_1 + A_2} \langle R_m(z) \rangle 
           \log (z - e_p^+) dz \right].
    \eea
  The second term is evaluated by deforming the contour to the cycle around the log cut as
    \bea
    \frac{1}{2 \pi i} \oint_{A_1 + A_2} \langle R_m(z) \rangle \log (z - e_p^+)
    &=&  - \int_{e_p^+}^{\infty} \langle R_m(z) \rangle dz
         + \frac{g_s N}{2 \pi i} \oint_{\infty} \frac{\log (z - e_p^+)}{z} dz
           \nonumber \\
    &=&  - \int_{e_p^+}^{\Lambda_0} \langle R_m(z) \rangle dz
         + g_s N (\log \Lambda_0 + \pi i) + \CO \left( \frac{1}{\Lambda_0} \right).
           \label{secondterm}
    \eea
  Note that $\langle R_m \rangle$ has a pole only at $z = \infty$.
  We regularized the integral by introducing the cut off $\Lambda_0$.
  We will take $\Lambda_0$ to infinity after the calculation.
  By changing $\langle R_m \rangle$ into $\lambda_m$ we obtain
    \bea
    \frac{\partial F_m}{\partial \nu_p}
     =   - 2 \sqrt{2} \pi i \int_{e_p^+}^{\Lambda_0} \lambda_m
         - \mu_0 \log \Lambda_0 + 2 \pi i g_s N + \CO \left( \frac{1}{\Lambda_0} \right).
           \label{Fderivative}
    \eea
    
  In order to compare with the prepotential of the gauge theory, we let $g_s \rightarrow g_s \epsilon$ as above, 
  which gives $\mu_i$, $\mu_0$, $\nu_p$ and $F_m$ dimension.
  Since $F_m$ has mass dimension 2, it is written as
    \bea
    2 F_m
     =     \sum_{i = 1}^3 \mu_i \frac{\partial F_m}{\partial \mu_i}
         + \sum_{p = 1,2} \nu_p \frac{\partial F_m}{\partial \nu_p}.
           \label{Fm}
    \eea
  The first term is evaluated as
    \bea
    \frac{\partial F_m}{\partial \mu_i}
    &=&    g_s \langle \Tr \log (M - q_i) \rangle
     =     \frac{1}{2 \pi i} \oint_{A_1 + A_2} \langle R_m(z) \rangle \log (z - q_i) dz
           \nonumber \\
    &=&    \frac{\pi i}{\sqrt{2}} \left[ \int_{\Lambda_0}^{q_i} + \int_{\tilde{q}_i}^{\tilde{\Lambda}_0} \right] \lambda_m
         - \frac{1}{2} W(q_i) - \frac{\mu_0}{2} \log \Lambda_0 + g_s N \pi i + \CO \left( \frac{1}{\Lambda_0} \right),
    \eea
  where we have deformed the contour to the cycle around the log cut as in (\ref{secondterm}).
  Also, by using (\ref{Fderivative}), the second term in (\ref{Fm}) is
    \bea
    \sum_{p = 1,2} \nu_p \frac{\partial F_m}{\partial \nu_p}
    &=&  - \sqrt{2} \pi i \left[ \nu_1 \int_{B_1} + \nu_2 \int_{B_2} \right] \lambda_m
         - g_s N \mu_0 \log \Lambda_0 + 2 \pi i (g_s N)^2 + \CO \left( \frac{1}{\Lambda_0} \right)
           \label{Fderivativenu} \\
    &=&    \frac{\pi i}{\sqrt{2}} \left( \sum_{i = 1}^3 \mu_i + \mu_0 \right) \int_{B_2} \lambda_m
         + \sqrt{2} \pi i \nu_1 \oint_B \lambda_m - g_s N \mu_0 \log \Lambda_0 + 2 \pi i (g_s N)^2
         + \CO \left( \frac{1}{\Lambda_0} \right),
           \nonumber
    \eea
  where $B_1$ and $B_2$ are the paths from $\tilde{\Lambda}_0$ to $\Lambda_0$ through the branch cuts respectively
  and we define the cycle $B = B_2 - B_1$.
  Combining these, we finally obtain
    \bea
    2 F_m
    &=&    \frac{\pi i}{\sqrt{2}} \left[ \sum_{i = 1}^3 \mu_i \int_{\tilde{q}_i}^{q_i} + \mu_0 \int_{B_2} \right] \lambda_m
         + \pi i \oint_{A_1} \lambda_m \oint_{B} \lambda_m
           \nonumber \\
    & &  - \frac{1}{2} \sum_{i = 1}^3 \mu_i W(q_i) + \frac{1}{2} \mu_0^2 \log \Lambda_0 - \pi i g_s N \mu_0 
         + \CO \left( \frac{1}{\Lambda_0} \right).
    \eea
  The first line of the above equation is the same as the prepotential of the gauge theory \cite{Matone, STY, EY}
  up to a factor $\pi i$ which can be absorbed by the redefinition of the free energy.
  Note that the divergence of integrals in the first line 
   is canceled by the divergence of  terms in the second line
  and the free energy is finite.
  There are also finite terms in the second line which depend only on the mass parameter $\mu_i$ and $\mu_0$.
  However, we should note that the terms in the first line have ambiguities due to the choice of integration paths.
  By deforming the paths, an additional contribution which is bilinear in the masses can appear.
  Therefore, we conclude that the free energy of the matrix model is the same as the prepotential
  up to these moduli independent terms.

\section{Matrix model for asymptotically free theories}
\label{sec:asymptoticfree}

\subsection{Asymptotically free theory with $N_f = 3$}
\label{subsec:asymptoticfree3}
  In this and next subsections, we consider the decoupling of heavy flavors from the matrix model (1.1) 
  and introduce the matrix theory with flavors $N_f<4$ which would describe the $\CN=2$ gauge theories 
  in the asymptotically free region.
  
  Before taking the decoupling limit we should recall the precise relationship 
  between $u$ and $\langle \Tr \phi^2 \rangle$ as pointed out in ref \cite{SeibergWitten2}
    \bea
    u
     = \langle \Tr \phi^2\rangle - \frac{1}{6} (\vartheta_4^4 + \vartheta_3^4) \sum_{i = 0}^3 m_i^2.
           \label{u}
    \eea
  By substituting (\ref{u}) into (\ref{solution}), we obtain
    \bea
    - (1 + q_{{\rm UV}}) U
     =   - \frac{\langle \Tr \phi^2\rangle}{\vartheta_3^4} + 2 q_{{\rm UV}} m_2 m_3
         + \frac{1}{2} q_{{\rm UV}} (m_3^2 - m_1^2) + \frac{1}{2} q_{{\rm UV}} (m_2^2 - m_0^2).
    \eea
  In the following subsections we again use the symbol $u$ to denote $\langle \Tr \phi^2\rangle$ for  simplicity. 
  We hope that no confusion should arise from this change of notation.
    
  In order to discuss the decoupling limit we write the mass parameters as follows
    \bea
    m_{\pm}
     =     m_2 \pm m_0,~~~
    \tilde{m}_{\pm}
     =     m_3 \pm m_1,
    \eea
  and consider the limit $\tilde{m}_- \rightarrow \infty, q_{{\rm UV}} \rightarrow 0$ 
  and the other masses and $\tilde{m}_- q_{{\rm UV}} = \Lambda_3$ fixed,
  where $\Lambda_{3}$ corresponds to the dynamical scale of the gauge theory.  
  In this case, the relation on the mass parameters (\ref{massrelation}) becomes
    \bea
    m_0 + m_2 + \tilde{m}_+
     =   - 2 g_s N,
           \label{massrelation'}
    \eea
  and this remains finite.
  The matrix model action leads to
    \bea
    W(M)
     =     \tilde{m}_+ \log M - \frac{\Lambda_{3}}{2 M} + m_2 \log (M - 1).
           \label{action3}
    \eea
  and we obtain the spectral curve for $N_f =3$ theory 
    \bea
    x^2
     =     \frac{\Lambda_{3}^2}{4 z^4} - \frac{\tilde{m}_+ \Lambda_3}{z^3 (z - 1)} 
         - \frac{u - (m_2 + \frac{1}{2} \tilde{m}_+) \Lambda_3}{z^2 (z - 1)}
         + \frac{m_0^2}{z (z - 1)} + \frac{m_2^2}{z (z - 1)^2} - \frac{m_2 \Lambda_3}{z^2 (z - 1)}.
           \label{McurveNf=3}
    \eea
  Note that the structure of the singularities is similar to that obtained in \cite{GMN}.
  Let us consider $u$-derivative of the Seiberg-Witten one-form $\lambda_{{\rm m}} = \frac{x dz}{2\sqrt{2} \pi i}$:
    \bea
    \omega
     =     \frac{\partial \lambda_{{\rm m}}}{\partial u}
     =     \frac{dz}{2 \sqrt{2} \pi i} \frac{-1}{2 x z^2 (z - 1) }
     \equiv
         - \frac{1}{4 \pi i \sqrt{2 Q_4(z)}},
    \eea
  where
    \bea
    Q_4(z)
    &=&    m_0^2 z^4 + \left(- u - m_0^2 + m_2^2 + \frac{1}{2} \tilde{m}_+ \Lambda_3 \right) z^3
           \nonumber \\
    & &  + \left( u + \frac{\Lambda_3^2}{4} - \frac{3}{2} \tilde{m}_+ \Lambda_3 \right) z^2
         + \left( - \frac{\Lambda_3^2}{2} + \tilde{m}_+ \Lambda_3 \right) z + \frac{\Lambda_3^2}{4}.
    \eea
  The discriminant of the polynomial above completely matches with that of the Seiberg-Witten curve for $N_f=3$ 
  in \cite{SeibergWitten2}:
    \bea
    y^2
    &=&    x^2 (x - u) - \frac{1}{4} \Lambda_3^2 (x - u)^2 
         - \frac{1}{4} (m_+^2 + m_-^2 + \tilde{m}_+^2) \Lambda_3^2 (x - u)
           \nonumber \\
    & &  + m_+ m_- \tilde{m}_+ \Lambda_3 x 
         - \frac{1}{4} (m_+^2 m_-^2 + m_-^2 \tilde{m}_+^2 + \tilde{m}_+^2 m_+^2) \Lambda_3^2.
    \label{N_f=3}
    \eea
  Note that the dynamical scale $\Lambda_3$ differs by a factor from $\Lambda_{{\rm SW}}$ that appeared in \cite{SeibergWitten2}.
  This is due to the difference of the decoupling limit of ours and that of \cite{SeibergWitten2} 
  where they fixed
    \bea
    64 \tilde{m}_- q_{{\rm IR}}
     \equiv 
           \Lambda_{{\rm SW}}.
    \eea
  As discussed above, $q_{{\rm UV}} = 16 q_{{\rm IR}}$ for the weak coupling region.
  Thus, we obtain $\Lambda_{{\rm SW}} = 4 \tilde{m}_- q_{{\rm UV}} = 4 \Lambda_3$.
  
  For the equal hypermultiplet mass case where $m_0 = 0$, $m_2 = m$ and $\tilde{m}_+ = m$, 
  the discriminant becomes
    \bea
    \Delta
    &=&    \frac{\Lambda_3^2}{32} \left( u - m^2 - \frac{1}{2} m \Lambda_3 \right)^3
           \left( -32 u^2 + 2 \Lambda_3^2 u - 48 u m \Lambda_3 + 3 m \Lambda_3^3
         + 6 m^2 \Lambda_3^2 + 128 m^3 \Lambda_3 \right).
           \nonumber \\
    & &
    \eea
  This implies that three ($SU(3)$ triplet) massless particles appear 
  at the singularity $u = m^2 + \frac{1}{2} m \Lambda_3$.
  This is the correct property of the curve (\ref{N_f=3}).

\subsubsection*{Free energy}
  Now we consider the free energy of this matrix model.
  The formula (\ref{Fderivative}) for the derivative of the free energy with respect to the filling fractions 
  can be used without change.
  By restoring the dimension, we obtain the following equation
    \bea
    2 F_m
     =     \left( \tilde{m}_+ \frac{\partial}{\partial \tilde{m}_+} + m_2 \frac{\partial}{\partial m_2}
         + \Lambda_{3} \frac{\partial}{\partial \Lambda_{3}} 
         + \sum_{p = 1,2} \nu_p \frac{\partial }{\partial \nu_p} \right) F_m.
           \label{differentialeq}
    \eea
  By using the result in the previous section, 
  the computation of the first two terms are straightforward: e.g., 
    \bea
    \frac{\partial F_m}{\partial \tilde{m}_+}
    &=&    \frac{\pi i}{\sqrt{2}} \left( \int_{\Lambda_0}^0 + \int_{\tilde{0}}^{\tilde{\Lambda_0}} \right) \lambda_m
         - \frac{1}{2} W(0) - \frac{m_0}{2} \log \Lambda_0 + \pi i g_s N + \CO \left( \frac{1}{\Lambda_0} \right).
    \eea
  Also, the last term is exactly the same as (\ref{Fderivativenu}) except that $\sum_{i = 1}^3 m_i$ becomes $\tilde{m}_+ + m_2$.
  Only the new ingredient which need a little bit care is the third term in (\ref{differentialeq}).
  In order to compute this, we note that in the spectral curve: $x^2 = W'(z)^2 + f(z)$, 
  $f(z)$ is written as 
    \bea
    f(z)
     =     \frac{f_1}{z} + \frac{f_2}{z - 1} + \frac{f_3}{z^2},
           ~~~
    f_3
     =   - 2 g_s \Lambda_3 \left< \sum_I \frac{1}{\lambda_I} \right>.
    \eea
  This $f_3$ can be directly evaluated from the curve (\ref{McurveNf=3}), which leads to
    \bea
    \frac{\partial F_m}{\partial \Lambda_{3}}
     =   - \frac{g_s}{2} \left< \sum_I \frac{1}{\lambda_I} \right>
     =     \frac{f_3}{4 \Lambda_{3}}
     =     \frac{1}{4 \Lambda_{3}} \left( u + ( \frac{1}{2} \tilde{m}_+ + m_2) \Lambda_{3} - \tilde{m}_+^2 \right).
    \eea
  
  By collecting these, we finally obtain the expression for the free energy
    \bea
    2 F_m
     =     \frac{\pi i}{\sqrt{2}} 
           \left( \tilde{m}_+ \int_{\tilde{0}}^0 + m_2 \int_{\tilde{1}}^1 + m_0 \int_{B_2} \right) \lambda_m
         + \pi i \oint_{A_1} \lambda_m \oint_B \lambda_m + \frac{u}{4} + \ldots.
    \eea
  This nicely matches with the prepotential of the theory with $N_f = 3$ \cite{Matone, STY, EY}
  up to the irrelevant factor $\pi i$.
  The ellipsis is the terms which depend only on the masses and the scale factor.
  It is reassuring to note that the beta function coefficient $b = 1$ in front of $u$ is
  correctly recovered in the matrix model computation.

\subsection{Asymptotically free theory with $N_f = 2$}
\label{subsec:asymptoticfree2}
  In this subsection, we propose a matrix model action which corresponds to $SU(2)$ gauge theory with $N_f = 2$.
  Method of its derivation is similar as the one in the previous section:
  we consider the limit where 
  $z$ is rescaled as $\frac{\Lambda_3}{\Lambda_2} z$ 
  and $m_- \rightarrow \infty$ while keeping $m_- \Lambda_{3} \equiv \Lambda_{2}^2$.
  The spectral curve is given by
    \bea
    x^2
     =     \frac{\Lambda_2^2}{4 z^4} + \frac{\tilde{m}_+ \Lambda_2}{z^3}
         + \frac{u}{z^2} + \frac{m_+ \Lambda_2}{z} + \frac{\Lambda_2^2}{4}.
           \label{SWcurve2}
    \eea
  Here, we have rescaled $x$ as $x \rightarrow \frac{\Lambda_2}{\Lambda_3} x$.
  This curve is the same form as the first realization of $SU(2)$ gauge theory with $N_f = 2$ in \cite{GMN}.
  In this limit, the action of the matrix model (\ref{action3}) becomes
    \bea
    W(M)
     =     \tilde{m}_+ \log M - \frac{\Lambda_{2}}{2 M} - \frac{\Lambda_{2} M}{2},
    \eea
  where we have ignored irrelevant constant terms.
  The relation on the mass parameters (\ref{massrelation'}) is, in this limit,
    \bea
    \tilde{m}_+ + m_+
     =   - 2 g_s N.
    \eea
    
  The holomorphic one-form is
    \bea
    \omega
     =     \frac{\partial}{\partial u} \frac{x dz}{2 \sqrt{2} \pi i}
     =     \frac{dz}{4 \sqrt{2} \pi i \sqrt{R_4(z)}},
    \eea
  where
    \bea
    R_4(z)
     =     \frac{\Lambda_2^2}{4} z^4 + m_+ \Lambda_2 z^3
         + u z^2 + \tilde{m}_+ \Lambda_2 z + \frac{\Lambda_2^2}{4}.
    \eea
  
  The discriminant of the above polynomial agrees completely with that of the Seiberg-Witten curve for $N_f=2$ 
  in \cite{SeibergWitten2}:
    \bea
    y^2
     =     (x^2 - \frac{1}{4} \Lambda_2^4)(x - u) + m_+ \tilde{m}_+ \Lambda_2^2 x 
         - \frac{1}{4} (m_+^2 + \tilde{m}_+^2) \Lambda_2^4.
    \eea
  
  For the equal mass case where $m_+ = \tilde{m}_+ = m$, the discriminant becomes
    \begin{eqnarray}
    \Delta
     =    \Lambda_2^4 
          \left( u + 2 m \Lambda_{N_f =2} + \frac{1}{2} \Lambda_2^2 \right) 
          \left( u - 2 m \Lambda_2 + \frac{1}{2} \Lambda_2^2 \right) 
          \left( u - m^2 - \frac{1}{2} \Lambda_2^2 \right)^2.
    \end{eqnarray}

\subsubsection*{Free energy}
  Let us compute the free energy of this model.
  The procedure is the same as that in section \ref{sec:matrix} and \ref{subsec:asymptoticfree3}.
  Only non-trivial point is the calculation of $\Lambda_{2}$ derivative:
    \bea
    \frac{\partial F_m}{\partial \Lambda_{2}}
     =   - \frac{g_s}{2} \left< \sum_{I} \left( \frac{1}{\lambda_I} + \lambda_I \right) \right>
     =     \frac{g_2}{4 \Lambda_{2}} - \frac{g_s}{2} \langle \sum_I \lambda_I \rangle,
    \eea
  where $g_2$ is the coefficient of $g(z) = \frac{g_1}{z} + \frac{g_2}{z^2}$
  in the spectral curve: $x^2 = W'(z)^{2} + g(z)$.
  By comparing this form of the spectral curve with (\ref{SWcurve2}), 
  we obtain $g_2 = u + \frac{1}{2} \Lambda_2^2 - \tilde{m}_+^2$.
  The second term is $\langle \sum_I \lambda_I \rangle = \langle \tr M \rangle$.
  This can be determined, by using the same argument as that in subsection \ref{subsec:recursion}, as
    \bea
    2 \Lambda_2 \langle \tr M \rangle
     =   - u - \frac{1}{2} \Lambda_2^2 + m_+^2.
    \eea
  Therefore, we obtain
    \bea
    \Lambda_{2} \frac{\partial F_m}{\partial \Lambda_{2}}
     =     \frac{1}{4} \left( g_2 + u + \frac{1}{2} \Lambda_2^2 - m_+^2 \right)
     =     \frac{u}{2} - \frac{m_+^2 + \tilde{m}_+^2 - \Lambda_2^2}{4}.
    \eea
  Then, the similar computation as above leads to
    \bea
    2 F_m
     =     \frac{\pi i}{\sqrt{2}} 
           \left( \tilde{m}_+ \int_{\tilde{0}}^0 + m_+ \int_{B_2} \right) \lambda_m
         + \pi i \oint_{A_1} \lambda_m \oint_B \lambda_m + \frac{2 u}{4} + \ldots.
    \eea
  This is the same form as the prepotential of the theory with $N_f = 2$.
  It is nice to find the appearance of the beta function coefficient $b = 2$ for $N_f=2$ theory in front of $u$ 
  from the matrix model computation.
  
\section{Conclusion and discussion}
 
 In this article we have studied the Penner type matrix model and gave strong evidence 
 that it reproduces correctly the physics of $\cal{N}=$2 supersymmetric $SU(2)$ gauge theories. 
 It will be interesting to see if the model can in fact be used to provide a 
 simple derivation of the AGT relation.
   
  Generalization to $SU(2)$ linear quiver gauge theories \cite{Gaiotto} may be straightforward.
  In this case, the corresponding action of the matrix model is
    \bea
    W(M)
     =     \sum_{i = 1}^{k-1} \mu_i \log(M - q_i),
    \eea
  where $k$ corresponds to the number of the simple poles of the Seiberg-Witten one-form.
  It will be interesting to study the S-duality group of this matrix theory. 
  
  A different type of matrix model has been proposed in ref \cite{KlemmSulkowski}
  to derive Seiberg-Witten theory. 
  It is important to see how this model is related to the matrix theory discussed in this paper.

\section*{Acknowledgements}
  We would like to thank K.~Hosomichi, Katsushi Ito, H.~Itoyama, A.~Marshakov, T.~Oota, N.~Sasakura and M.~Taki 
  for discussions. 
  Research of T.E. is supported in part by Grant in Aid from the Japan Ministry of
  Education, Culture, Sports, Science and Technology.
  K.~M.~ is supported in part by JSPS Research Fellowships for Young Scientists.




\end{document}